\documentclass[aps,twocolumn,showpacs,amsmath,amssymb,superscriptaddress,final,prb]{revtex4-1}
\usepackage[latin1]{inputenc}   
\usepackage{graphicx}  
\usepackage{dcolumn}  
\usepackage{nicefrac}   
\graphicspath{{./}{figure/}}
\usepackage[dvips]{color} 
\definecolor{red}{rgb}{0.85,.1,0}
\definecolor{cblue}{named}{CadetBlue}

\newcommand{\ypd}{Y\-Pd$_{2}$\-Sn }
\newcommand{\mSR}{$\mu$SR }
\newcommand{\Tc}{$T_{c}$ }
\newcommand{\scg}{superconducting }

\renewcommand{\vec}[1]{\boldsymbol #1}

\begin{document}
\title{$\mu$SR and NMR study of the superconducting Heusler compound YPd$_2$Sn} %

\author{H.~Saadaoui}
\affiliation{Laboratory for Muon Spin Spectroscopy, Paul Scherrer Institut, CH-5232 Villigen PSI, Switzerland}
\author{T.~Shiroka}
\affiliation{Laboratorium f\"ur Festk\"orperphysik, ETH-H\"onggerberg, CH-8093 Z\"urich, Switzerland}
\affiliation{Paul Scherrer Institut, CH-5232 Villigen PSI, Switzerland}
\author{A.~Amato}
\affiliation{Laboratory for Muon Spin Spectroscopy, Paul Scherrer Institut, CH-5232 Villigen PSI, Switzerland}
\author{C.~Baines}
\affiliation{Laboratory for Muon Spin Spectroscopy, Paul Scherrer Institut, CH-5232 Villigen PSI, Switzerland}
\author{H.~Luetkens}
\affiliation{Laboratory for Muon Spin Spectroscopy, Paul Scherrer Institut, CH-5232 Villigen PSI, Switzerland}
\author{E.~Pomjakushina}
\affiliation{Laboratory for Developments and Methods, Paul Scherrer Institut, CH-5232 Villigen PSI, Switzerland}
\author{V.~Pomjakushin}
\affiliation{Laboratory for Neutron Scattering, Paul Scherrer Institut, CH-5232 Villigen PSI, Switzerland}
\author{J.~Mesot}
\affiliation{Laboratorium f\"ur Festk\"orperphysik, ETH-H\"onggerberg, CH-8093 Z\"urich, Switzerland}
\affiliation{Paul Scherrer Institut, CH-5232 Villigen PSI, Switzerland}
\author{M.~Pikulski}
\affiliation{Laboratorium f\"ur Festk\"orperphysik, ETH-H\"onggerberg, CH-8093 Z\"urich, Switzerland}
\affiliation{Paul Scherrer Institut, CH-5232 Villigen PSI, Switzerland}

\author{E.~Morenzoni}
\email[E-Mail: ]{elvezio.morenzoni@psi.ch} \affiliation{Laboratory for Muon Spin Spectroscopy, Paul Scherrer Institut, CH-5232 Villigen PSI, Switzerland}
\date{\today}

\begin{abstract}
We report on muon spin rotation/relaxation and $^{119}$Sn nuclear magnetic resonance (NMR)
measurements to study the microscopic superconducting and magnetic properties of the Heusler compound with the highest superconducting transition temperature, \ypd\ ($T_c=5.4$ K).
Measurements in the vortex state provide the temperature dependence of the effective magnetic penetration depth $\lambda(T)$ and the field dependence of the superconducting gap $\Delta(0)$. The results are consistent with a very dirty $s$-wave BCS superconductor with a gap $\Delta(0)=0.85(3)$ meV, $\lambda(0)= 212(1)$ nm, and a Ginzburg-Landau coherence length $\xi_{\mathrm{GL}}(0)\cong 23$ nm. In spite of its very dirty character, the effective density of condensed charge carriers is high compared to the normal state. The \mSR data in a broad range of applied fields are well reproduced by taking into account a field-related reduction of the effective superconducting gap. Zero-field \mSR measurements, sensitive to the possible presence of very small magnetic moments, do not show any indications of magnetism in this compound.
\end{abstract}
\maketitle

\section{\label{Sec:intro}Introduction}
Heusler materials \cite{heusler03} display a large variety of
interesting electronic properties such as different types of magnetic order
(ferro-, antiferro-, and ferrimagnetism), heavy fermion behavior, half metallic
ferromagnetism, and superconductivity. \cite{graf11} Some materials of this class may also exhibit properties typical of a topological
insulator.\cite{butch11} Their electronic tunability and multifunctionality make them attractive candidates for spintronics applications. The so-called full Heusler compounds with the general formula AT$_2$M crystallize in the cubic $L21$ structure corresponding to four interpenetrating fcc sublattices. The majority of these phases contain a main group element (M), a transition element (T), and either a rare earth or another transition element (A).
Out of many hundreds of Heusler compounds only less than 30 (with Pd, Ni, or Au at the T site) are superconducting at ambient pressure.

Recent transport and thermodynamic studies have identified strong electron-phonon coupling as the most important factor leading to superconductivity in these families. \cite{klimczuk12} However, in contrast to the simple BCS theory, no conventional dependence of $T_c$ as a function of the BCS parameters, such as $N(0)$ or the Debye temperature, was found.\cite{klimczuk12}
Among the superconducting Heusler compounds, those based on Pd (APd$_2$M) are of special interest because they form the largest family.
Here \Tc shows a pronounced dependence on $N(0)$; an increase by only $\sim$20\% of $N(0)$ more than doubles $T_c$, indicating the importance of electronic instabilities. Like most of the \scg\ Heusler compounds, \ypd\ has 27 valence electrons, implying a ratio of valence electrons/atom close to 7 (as required by the Matthias rule \cite{matthias53}) and exhibits the highest \Tc\ among all known superconducting Heusler alloys.\cite{ishikawa82}
In similar Pd-based compounds, the occurrence of superconductivity has been predicted using electronic structure calculations
and confirmed experimentally. Their band structures exhibit saddle points at the L point close to the Fermi energy $E_{\mathrm{F}}$, resulting in a high density of states (DOS) and leading to a so-called Van Hove singularity in the energy-dispersion
curve.\cite{winterlik09}

Unlike conventional BCS superconductors, several Pd-based compounds, such as YbPd$_2$Sn \cite{kierstead85} and ErPd$_2$Sn,\cite{shelton86} display co-existing superconductivity and long-range antiferromagnetic order. A clear and complete understanding of the origin of superconductivity, magnetism, and especially of their coexistence or interplay in the full Heusler compounds, is still missing. In YbPd$_2$Sn, earlier \mSR\ and inelastic neutron scattering measurements indicated an interplay between the superconducting state and magnetic correlations,\cite{amato03}
with possible weak magnetic coupling between the Pd and Yb-Sn sublattices.\cite{aoki00} In Y\-Pd$_{2}$\-Sn, earlier \mSR data also showed an anomaly of possible magnetic origin setting in at $T_c$.\cite{amato04}
For a non-magnetic superconductor such as \ypd\, the large variation of \Tc\!'s reported in the literature is also striking.  In this respect it is of interest to characterize the magnetic and superconducting properties of Y\-Pd$_2$\-Sn at a microscopic level and to clarify the role of Pd, which due to its large Stoner factor is near to ferromagnetism.\cite{chen89}

Here, we investigated the local superconducting and magnetic properties of \ypd\  by \mSR\ and NMR techniques, which are sensitive spin probes to weak magnetism of static or dynamic origin.
In particular, we studied the $B$-$T$ phase diagram in the vortex state in a magnetic field over a large range of fields to determine, besides $H_{c2}(T)$, the characteristic length
scales of  the superconductor, such as the effective magnetic penetration depth $\lambda(T)$ and the coherence lengths $\xi_0$ and $\xi_{\mathrm{GL}}(0)$.
The temperature dependence of $\lambda$ provides information about the superconducting gap parameters and symmetry. To search for evidence of possible weak magnetism in the normal and superconducting state, some measurements were performed in zero-field (ZF).

The paper is organized as follows: sample preparation and experimental details are described in Sec.\ II. In Sec.\ III we present results and details of the \mSR and NMR experiments in the vortex state and in ZF and discuss the normal and superconducting state properties and parameters. Conclusions follow in Sec.\ IV.

\section{\label{Sec:exp}Sample preparation and experimental details}
\begin{figure}[t]
\centering
\includegraphics[width=\columnwidth]{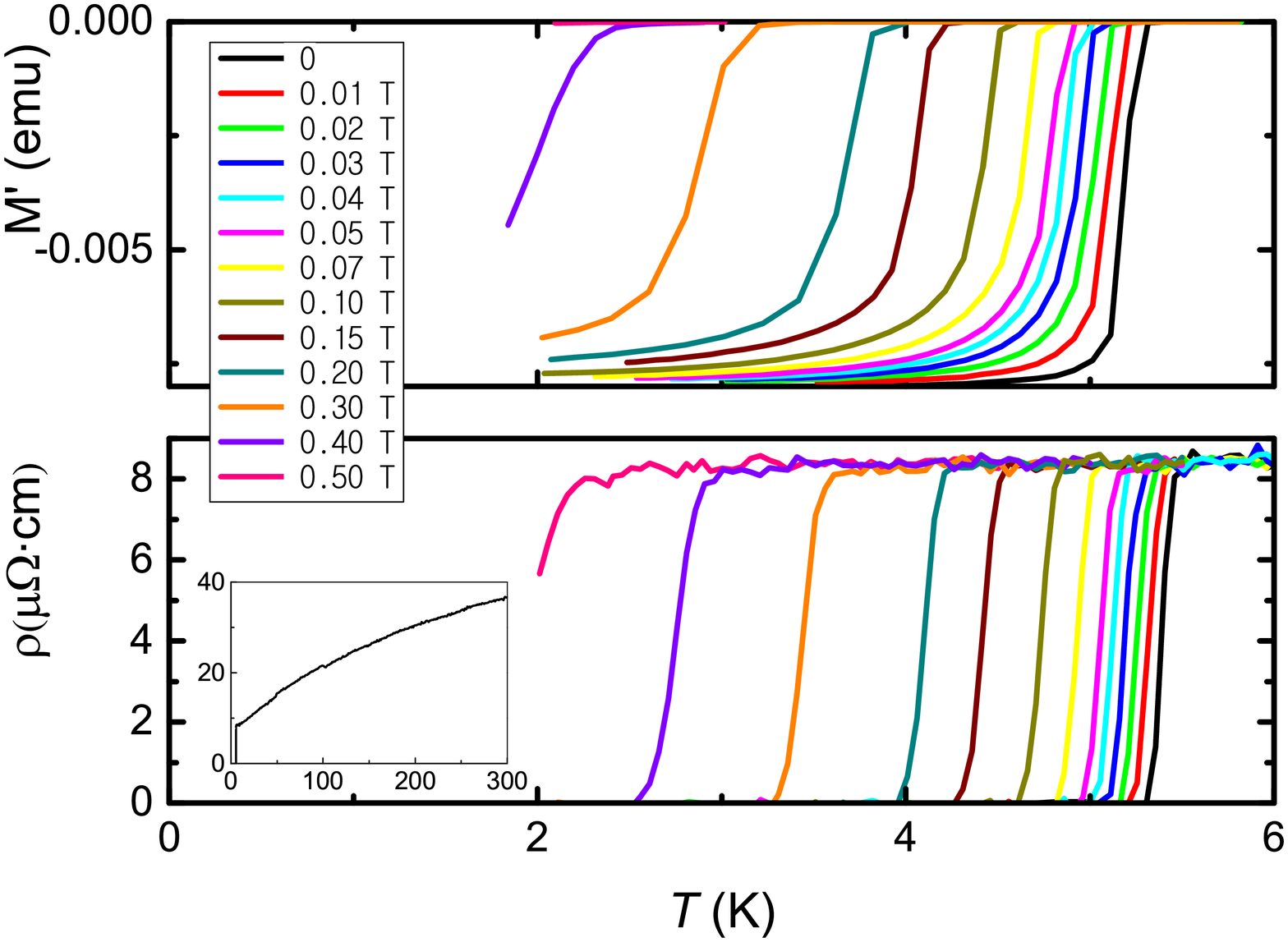}
\caption{(Color online) (Top) Low-temperature AC magnetization of Y\-Pd$_2$\-Sn in different applied dc fields using a 0.01 mT $H_{\mathrm{ac}}$ excitation at 1 kHz. (Bottom) Resistivity close to $T_c$ measured at different applied magnetic fields. The inset shows the resistivity over the full temperature range measured in zero field.}
\label{Fig:AC-res}
\end{figure}

A polycrystalline sample of \ypd was synthesized by the arc melting of the respective amounts of the elements; 4\% excess of tin was taken to compensate its loss during melting. Afterwards the sample was sealed into the evacuated quartz ampoule and annealed at $750^{\circ}\mathrm{C}$ for 180h. For the \mSR measurements we used a single, as-grown piece of about $4 \times 3 \times 2$ mm$^3$ in size. For the NMR measurements the sample was powderized and mixed with an inert oxide.
AC magnetic susceptibility (field-cooled and zero-field cooled) was measured with a Quantum Design PPMS. Characterization by powder x-ray diffraction at room temperature, with a D8 Advance Bruker AXS diffractometer using Cu K${\alpha}$ radiation ($\lambda = 0.1546$ nm),
and by neutron powder diffraction at the SINQ spallation source of the Paul
Scherrer Institute using the high resolution diffractometer for thermal neutrons HRPT (at low temperatures) confirmed the good single-phase quality of the sample and found lattice parameters consistent with literature.
Figure~\ref{Fig:AC-res} shows the low-temperature magnetization under zero and applied fields up to 0.5 T and the corresponding resistivity measurements. From the magnetization curves $M=M(T)$, we could determine the $B_{c2}(T)$
boundary, defined as the magnetic field where the sample enters the normal state (at a given temperature). In the resistivity curves we used the temperature value where
$\rho$ has dropped to half of its normal state value.
In absence of an applied field, we obtain $T_c=5.4\pm 0.05$ K, consistent with the transition from the normal
to the vortex state ($T=5.35$ K), as determined from $\mu$SR data extrapolated to zero field.
Finally, the value defined by the 50\ \% drop of ZF AC-susceptibility is $T_c=5.2$ K (with a 10--90\ \% width of $\pm 0.25$ K).

The \mSR\ experiment was performed at the GPS spectrometer at the $\pi$M3 beam line of the S$\mu$S facility at the Paul Scherrer Institute. This instrument covers temperatures down to 1.6 K and fields up to 0.6 T.
Low-temperature measurements (down to 0.05 K) were carried out at the LTF instrument at the same beam line.  In transverse-field (TF) experiments a field between 5 mT and 0.5 T is applied above the \Tc before gradually cooling down to base temperature and acquiring the data, thereby ensuring that the sample is in the vortex state.
In ZF experiments, the sample is cooled to 1.6 K and the \mSR\ spectra are collected upon warming up.
The stray fields at the sample position are compensated to the $\mu$T level by means of active, three-component field compensation, with the three pairs of correction coils being controlled by a flux-gate magnetometer. Forward and backward positron detectors (relative to the $\sim$100\% polarization of the incident muon beam) were used for the detection of the \mSR asymmetry time spectrum $A(t)$, which is proportional to the time evolution of the polarization of the muon ensemble stopped in the sample.

The nuclear magnetic resonance (NMR) technique provides complementary information to the interstitial-site picture of $\mu$SR.  \cite{maclaughlin76,rigamonti98,walstedt08} All constituent elements of YPd$_2$Sn contain NMR-active nuclei: the spin-$\nicefrac{1}{2}$ ${}^{119}$Sn and ${}^{89}$Y, and the spin-$\nicefrac{5}{2}$ ${}^{105}$Pd. While all of them can be used in the normal state, only ${}^{119}$Sn is suitable for investigations in the superconducting state of YPd$_2$Sn. The low value of the upper critical field $\mu_0 H_{c2} \simeq$ 0.7 T and the decrease of $T_c$ with field pose stringent limits to the highest field one can apply during temperature scans. By setting, e.g., $\mu_0 H_{\mathrm{ext}} = 0.4$ T one can access the superconducting state only below $\sim$ 3 K. At such low field the Larmor frequencies of ${}^{89}$Y and ${}^{105}$Pd are both below 1 MHz, which makes them undetectable by standard NMR techniques. As a result, for our NMR investigations of the superconducting state, only the ${}^{119}$Sn nuclei can be used.

\section{\label{Sec:results}Results and discussion}
\subsection{\label{Ssec:msr spectra nmr lines} TF-\mSR spectra and NMR lineshapes}

Muon spin rotation (\mSR\!) spectra in the vortex state as a function of temperature were taken by field-cooling the sample at different fields.  Different
models based on the London or Ginzburg-Landau theory and empirical approximations using multiple Gaussian are available to analyze the data and determine, from the inhomogeneous field distribution in the vortex state, the characteristic length scales of type-II superconductors, such as the magnetic penetration depth and the coherence length.\cite{yaouanc11,sonier00}
The limits of validity and reliability of these approaches have been investigated in detail in Ref.~[\onlinecite{maisuradze09}]. In agreement with
the latter, we find that the best fit over the entire range of applied fields is obtained using a multi-component Gaussian curve to represent the field distribution. Correspondingly, after subtraction of the background contribution, we fitted the time spectra with following expression using the musrfit package:\cite{suter12}
\begin{eqnarray}
A(t)=e^{-\frac{\sigma_{n}^2 t^2}{2}}\sum_{i=1}^{N}A_{i}e^\frac{-\sigma_{i}^2 t^2}{2}\cos{(\gamma_{\mu}B_{i}t+\varphi)}.
\label{Asy}
\end{eqnarray}
Here $\varphi$ is the initial muon spin phase, while $A_i$, $\sigma_{i}$, and $B_i$ are the amplitude, relaxation rate and first moment of the internal field of the $i$-th Gaussian component, respectively. $\sigma_{n}$ is a (small) contribution to the field distribution arising from the nuclear moments. It is independent of temperature and was determined well above $T_c$.

The second moment of the multi-Gaussian internal field distribution is then given by \cite{weber93}:

\begin{eqnarray}
\langle\Delta B^2\rangle=\frac{\sigma_{s}^2}{\gamma_{\mu}^2} = \sum_{i=1}^{N} && \, \frac{A_i}{A_1+A_2+\ldots+A_N} \nonumber \\
&&\, \left[(\frac{\sigma_{i}}{\gamma_{\mu}})^{2}+(B_i-\langle B \rangle)^2 \right],
\label{Secondmoment}
\end{eqnarray}
with
\begin{eqnarray}
\langle B\rangle=\sum_{i=1}^{N}\frac{A_i B_i}{A_1+A_2+\ldots+A_N}.
\label{Firstmoment}
\end{eqnarray}
The number of components $N$ is increased until the $\chi^2$ of the fit does not change significantly. Typically, we found that $N$ is between 1 (high-field data) and 3 (low-field data).

\begin{figure}[tbh]
\centering
\includegraphics[width=0.42\textwidth]{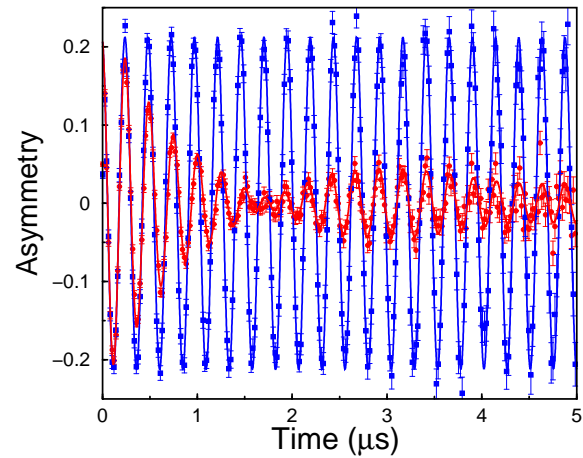}
\caption{\label{Fig:msr spectra}(Color online) Asymmetry spectrum in a TF of 30 mT. Blue symbols (squares): normal state. Red symbols (circles): 2 K data (field cooled), showing the typical damping resulting from the formation of a vortex lattice. Lines are fits using Eq.~\eqref{Asy}.}
\end{figure}

Figure~\ref{Fig:msr spectra}  shows the $\mu$SR spectra in a TF of 30 mT above and below $T_c$, fitted with Eq.~\eqref{Asy}. The spectrum at low temperature  clearly shows a pronounced damping reflecting the field inhomogeneity associated with the formation of
a flux-line lattice (FLL). The amplitude of this signal reflects a full-volume superconducting fraction (a background signal lower than 10\%  is due to muons missing the sample). Moreover, the non-Gaussian damping indicates the formation of an asymmetric field distribution, typical of a regular flux-line lattice.

\begin{figure}[t]
\centering
\includegraphics[width=0.42\textwidth]{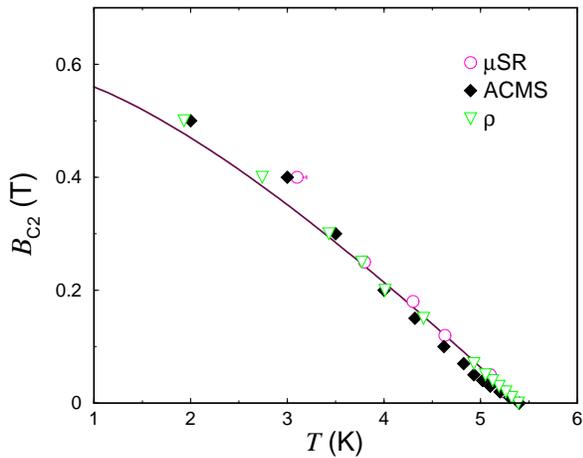}
\caption{\label{Fig:Bc2-vs-T}(Color online) Temperature dependence of the upper critical field of Y\-Pd$_2$\-Sn as obtained from resistivity measurements (triangles), $\mu$SR measurements of the temperature dependence of $\sigma_s$ (circles), and the complete disappearance of the AC magnetization in field (diamonds) (see Fig.~\ref{Fig:AC-res}b).
The solid line is the prediction of the WHH model.}
\end{figure}

A similar procedure was employed in the NMR experiments during the
{$^{119}$Sn lineshape measurements. The lowest applied field was
$\mu_0 H_{\mathrm{ext}}=0.224$ T, corresponding to a $^{119}$Sn NMR frequency
of 3.563 MHz. The NMR spectra exhibit rather narrow Gaussian lines in the normal state. Upon cooling below \Tc one notices a broadening of the lineshape as expected in a vortex solid. Figure \ref{Fig:NMR_line} illustrates this by reporting the $^{119}$Sn NMR lineshapes, taken at a field of 0.463 T, above and below $T_c$.

\begin{figure}[h]
\centering
\includegraphics[width=0.42\textwidth]{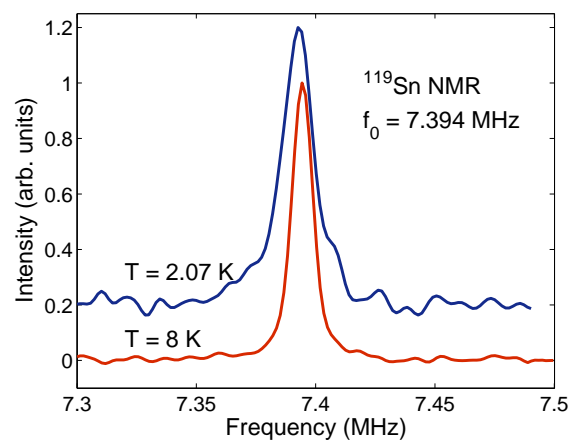}
\caption{\label{Fig:NMR_line}(Color online) ${}^{119}$Sn NMR lineshapes taken at 2.7
and at 8 K in $\mu_0 H_{\mathrm{ext}} = 0.463$ T. Both lines peak at nearly the same frequency, but the linewidth increases as the temperature is decreased.}
\end{figure}
\subsection{\label{Ssec:field distribution} Field distribution in the vortex state}

Figure~\ref{Fig:sigma-vs-T-B} shows the temperature dependence of the depolarization rate $\sigma$
as obtained from $\mu$SR measurements in different applied transverse fields.

Typically $\sigma$ increases from $\sigma_{\mathrm n}$ in the normal state,
to $\sigma=\sqrt{\sigma^2_{\mathrm{s}} + \sigma^2_{\rm n}}$ below $T_c$,
reflecting the inhomogeneous field distribution in the vortex state due to the
decrease of the effective magnetic penetration depth $\lambda$ or, equivalently, to the increase of the superfluid density $n_s$ when entering the vortex state.
Correspondingly, a gradual increase of the NMR linewidth as the temperature decreases is observed. It is interesting to note that the magnitude of the field broadening obtained by the two techniques agrees fairly well. This can be seen in Fig.~\ref{Fig:sigma-vs-T-B}, where the NMR line width is scaled to the muon data by taking into account the different gyromagnetic ratios
$\gamma_{^{119}\mathrm{Sn}} / \gamma_{\mu}$ = 0.1178.
However, the NMR data seem to be affected by additional broadening effects, probably due to the proximity to the phase boundary.
This reflects the limitations of NMR with respect to $\mu$SR when studying the vortex state of superconductors with low $B_{c2}$ values.

Figure~\ref{Fig:sigma-vs-T-B} also shows that $\sigma_s(T)$ decreases
as the field increases toward $B_{c2}$. This reflects the narrowing
of the magnetic field distribution as the inter-vortex distance decreases with field
and the contribution of the vortex cores, where the superconducting order parameter is suppressed, increases.
\begin{figure}[tbh]
\centering
\includegraphics[width=0.42\textwidth]{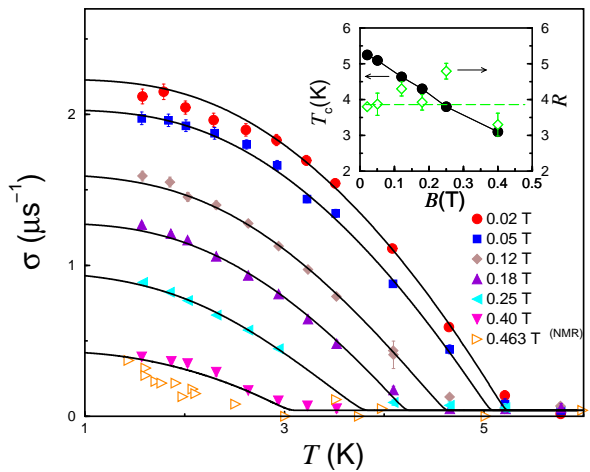}
\caption{\label{Fig:sigma-vs-T-B}(Color online) Temperature dependence of the muon spin depolarization rate in the vortex state of Y\-Pd$_2$\-Sn, measured after cooling in different applied fields. Also plotted are the NMR linewidths measured at 0.463 T and converted into equivalent $\mu$SR relaxation rates. The fit lines were obtained using the procedure explained in the text. Inset: $T_c$ and the gap-to-\Tc ratio $R \equiv 2 \Delta(0) / (k_{\mathrm{B}} T_c)$ vs.\ the applied magnetic field.}  
\end{figure}

The effect of the vortex cores and the expected field dependence of the second moment
of field distribution have been calculated using different models.\cite{brandt88, brandt03,yaouanc97}
One finds that the standard deviation in an applied field can be written as $\sigma_s (T,b)=\sigma_{s,L} f(b)$, where
\begin{equation}
\sigma_{s,L}(T)[\mu\mathrm{s}^{-1}]=\frac{1.0728 \times 10^5}{\lambda^2(T)[\mathrm{nm}^2]}
\end{equation}
is the expression obtained in the London limit,
with $b\equiv \langle B \rangle / B_{c2}$. From a calculation based on the Ginzburg-Landau model \cite{brandt03} and one based on the London model with a Gaussian cutoff (to take into account the finite size of vortex cores),\cite{brandt88}  analytical forms of $f(b)$ have been calculated:
\begin{align*}
f(b)&=0.452 (1-b)\left[1+1.21(1-\sqrt{b})^3\right] \qquad {\text{(GL model)}},\\
f(b)&=0.452 (1-b) \sqrt{1+3.9(1-b^2)} \quad {\text{(London with cutoff)}}.
\end{align*}
These forms are similar and differ only in high-order corrections to the linear field dependence.

\begin{figure}[t]
\centering
\includegraphics[width=0.42\textwidth]{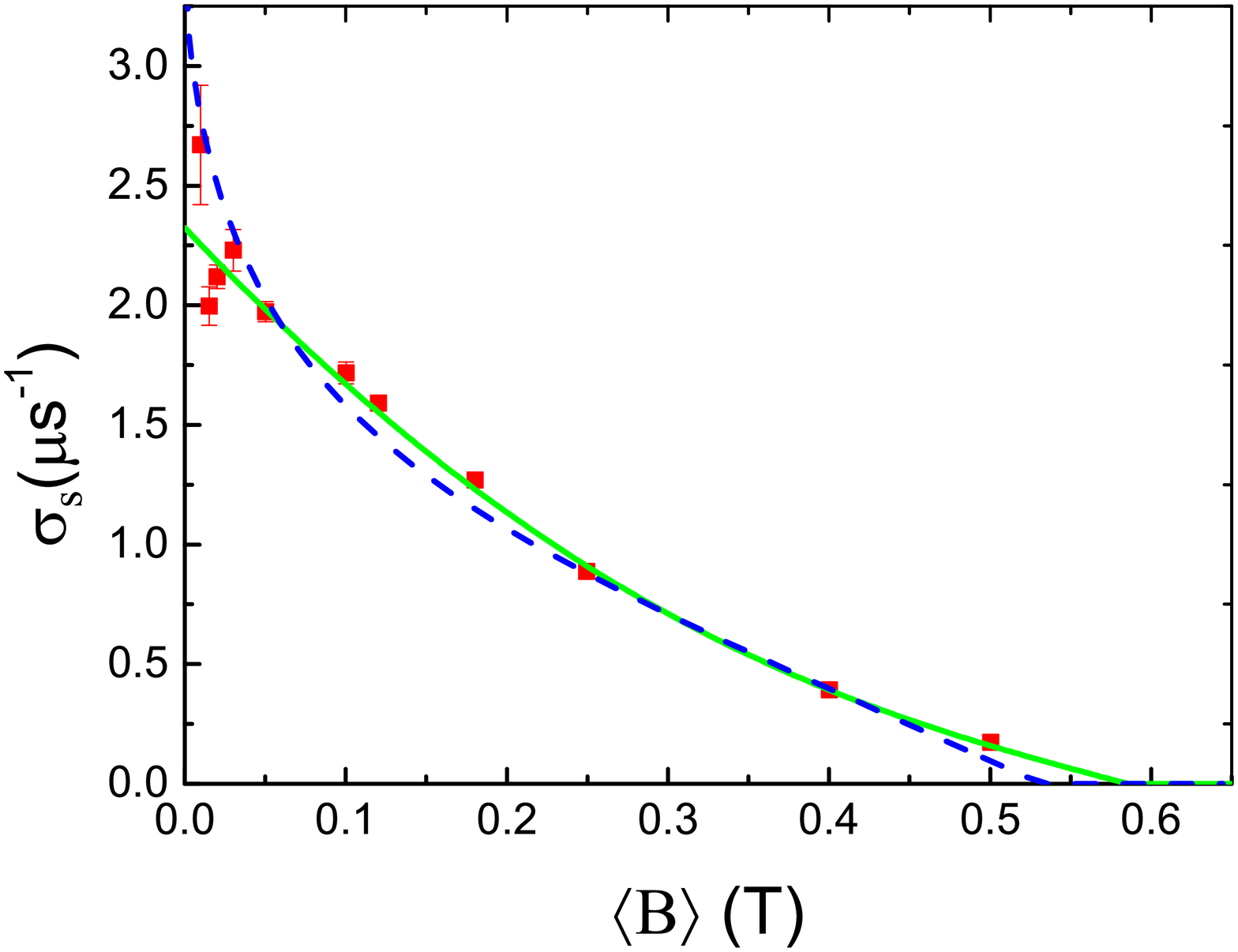}
\caption{\label{Fig:sigma-vs-b}(Color online) Field dependence of $\sigma_s$ measured
at $T=1.6$ K. Fits correspond to the modified London model\cite{brandt88} (solid line) and to the Ginzburg-Landau model\cite{brandt03} (dashed line).}
\end{figure}
It should be noted that the relevant parameter, which determines the characteristics of the vortex state is not $\mu_{0} H_{\mathrm{ext}}$,
but rather ${\langle B \rangle}$,} the average magnetic induction.
For $\mu_0 H_{\mathrm{ext}} \gg B_{c1}$ the magnetization $M$ is small, so that one
could safely use $\mu_0 H_{\mathrm{ext}}$ instead of $\langle B \rangle$. This does not hold at low fields, where also the demagnetization effects, which depend
on the sample shape, play a role. As a local probe technique, $\mu$SR provides the necessary parameter $\langle B \rangle$, thus avoiding the need for corrections. Figure~\ref{Fig:sigma-vs-b} shows the field dependence of $\sigma_s$ at 1.6 K. Both models were fitted to the data with $\lambda$ and $B_{c2}$ as free parameters. The equation based on the modified London model seems to reproduce better the field dependence
of $\sigma_s$ and $B_{c2}$ at 1.6 K (see Fig.~\ref{Fig:Bc2-vs-T}, where we show the $B_{c2}$ values as obtained
from AC susceptibility, resistivity, and the temperature dependence of $\sigma_s$ at different applied fields).
In the latter case we identify $B_{c2}(T)$ with $\mu_0 H_{\mathrm{ext}}$
at the temperature where $\sigma_s(T, H_{\mathrm{ext}})=0$.
The data points obtained by different methods are in good agreement and are well reproduced by the prediction of the Werthamer-Helfand-Hohenberg (WHH)\cite{werthamer66} expression for $B_{c2}(T)$.
The WHH model contains two parameters: the Maki parameter $\alpha$, which is a measure of the importance of the Pauli paramagnetism, and $\lambda_{so}$, the spin-orbit scattering constant. We have determined both parameters from our data:
$\alpha = 0.084$ and $\lambda_{so} = 1.5$ (see Sec.~\ref{Ssec:parameters}). The low value of the Maki parameter implies that the critical field due to Pauli paramagnetism $\mu_0 H_{\mathrm{P}} = \Delta(0) / (\sqrt{2}\mu_{\mathrm{B}}) \simeq 10$ T, is much higher than $B_{c2}(0)$, so that superconductivity is only limited by orbital currents.
Close to \Tc the temperature dependence of $B_{c2}$ is linear with a slope $\frac{dB_{c2}}{dT}=-0.16$ T/K. Using the well-known WHH formula  $B_{c2}(0)=-0.693\, T_c\, \frac{dB_{c2}}{dT}$ one can estimate $B_{c2}(0)$ = 0.6 T, which is very close to 0.57 T, the value extrapolated from our $B_{c2}(T)$ data.

The temperature dependence of the magnetic penetration depth is a measure of the superfluid density $n_s \propto 1 / \lambda^2$. In our case the $\mu$SR data can be well fitted using the expression for a dirty superconductor with a single $s$-wave gap:\cite{tinkham04}
\begin{eqnarray}
\frac{n_s(T)}{n_s(0)}=\frac{\lambda^2(0)}{\lambda^2(T)}= \frac{\Delta(T,B)}{\Delta(0,B)}\,\tanh\frac{\Delta(T,B)}{2 k_{\mathrm{B}} T},
\label{rho-vs-T}
\end{eqnarray}
\begin{figure}[th]
\centering
\includegraphics[width=0.42\textwidth]{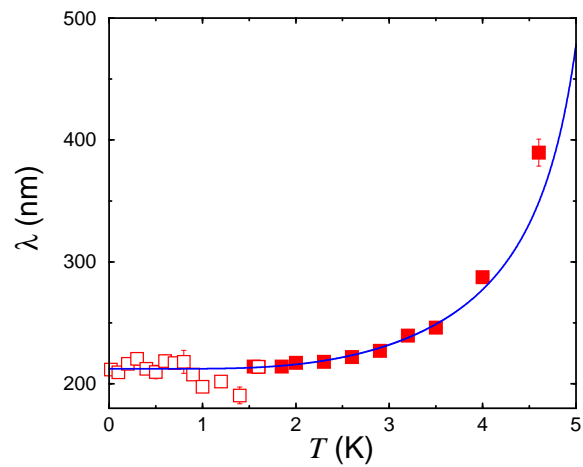}
\caption{\label{Fig:lambda-vs-T}(Color online) Temperature dependence of the effective magnetic penetration depth $\lambda$, fitted by assuming an $s$-wave order parameter in the dirty limit.}
\end{figure}
where $\Delta(T,B)$ is the BCS superconducting gap.
The dirty character of Y\-Pd$_2$\-Sn is confirmed by our determination of the normal state and superconducting parameters, as discussed in Sec.~\ref{Ssec:parameters}.
A useful parametrization of the BCS gap is given by Gross {\em et al.}\cite{gross86}
\begin{equation}
\Delta(T,B)=\Delta(0,B) \tanh\left[ \frac{\pi k_{\mathrm{B}} T_c(B)}{\Delta(0,B)} \sqrt{\frac{T_c(B)}{T}-1}\;\right].
\end{equation}
A good global fit of the \mSR data in Fig.~\ref{Fig:sigma-vs-T-B} is obtained by taking into account the effect of the applied field on the critical temperature and the superconducting gap. Fields of magnitude not negligible with respect to $B_{c2}$ act as pair breakers, smear out the sharp edges of the spectroscopic gap leading to an effective reduction of $\Delta(0)$. To take this into account, we have fitted our data allowing
not only for a field dependent $T_c$, but also for a field-dependent gap, $\Delta(0,B)$.
The results of the fits are shown in Fig.~\ref{Fig:sigma-vs-T-B} and
its inset. Interestingly, the BCS ratio $R \equiv 2 \Delta(0) / (k_{\mathrm{B}} T_c)$
appears to a good approximation to be field-independent, with a value at low field
of 3.85(9). From the usual approximation $B_{c2}(T)= B_{c2}(0)(1- (\frac{T}{T_c})^2)$, one obtains $T_{c}(B)= T_{c}(0)\sqrt{(1- \frac{B}{B_{c2}(0)}}$. Therefore, a constant gap to \Tc ratio implies a field dependence of the gap $\Delta(T=0,B)=\Delta(T=0,B=0)\sqrt{1 -  \frac{B}{B_{c2}(0)}}$. Such a field dependence is in agreement with the expected proportionality between the BCS gap and the spatially averaged Ginzburg-Landau order parameter and its field dependence $\Delta(T=0,B)^{2} \propto \langle | \Psi_{B}(\vec{r}) |^{2} \rangle = \Psi_{0}^{2} (1 -  \frac{B}{B_{c2}(0)})$.

The temperature dependence of $\lambda(T)$ is shown in Fig.~\ref{Fig:lambda-vs-T}, where we also plot the low-temperature data
obtained at 20 mT at the LTF instrument. The background signal at the LTF instrument is significantly higher than at the GPS instrument. Therefore the absolute value of $\lambda(T)$ is more precisely determined from the GPS than from the LTF data. For this reason we normalized the data at the common point of 1.6 K.
At low temperatures, there is no pronounced temperature dependence, as expected for a fully gapped $s$-wave superconductor, where $\Delta \lambda(T)= \lambda(T) - \lambda(0)$ decays exponentially. From the curve we extract a value $\lambda(0)$= 212(1) nm which, as expected for an $s$-wave superconductor, is found to be field independent.

\subsection{\label{Ssec:NMR shift} NMR relaxation and shift}
\begin{figure}[tbh]
\centering
\includegraphics[width=0.42\textwidth]{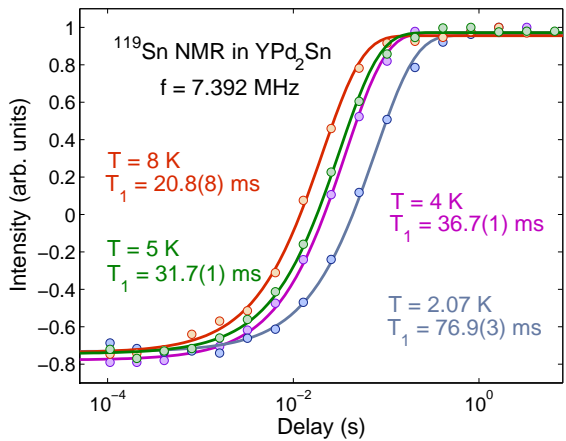} 
\caption{\label{Fig:NMR-relax}(Color online) Recovery of ${}^{119}$Sn 
longitudinal magnetization in YPd$_2$Sn at selected temperatures. The full lines refer to simple exponential fits of the spin-lattice relaxation corresponding to the reported $T_1$ values.}
\end{figure}

NMR spin-lattice relaxation-rate measurements are sensitive to dynamic electronic state properties. In our case we used a standard inversion-recovery technique realized by means of a Hahn spin-echo sequence, with a variable delay between the inversion pulse and the echo detection. Given the low temperature range, the inversion technique is preferred to the (faster) saturation recovery, since it involves a smaller number of RF pulses and, hence, avoids sample heating. The results of $T_1$ measurements for a selected number of temperatures are shown in Fig.~\ref{Fig:NMR-relax}.

\begin{figure}[tbh]
\centering
\includegraphics[width=0.42\textwidth]{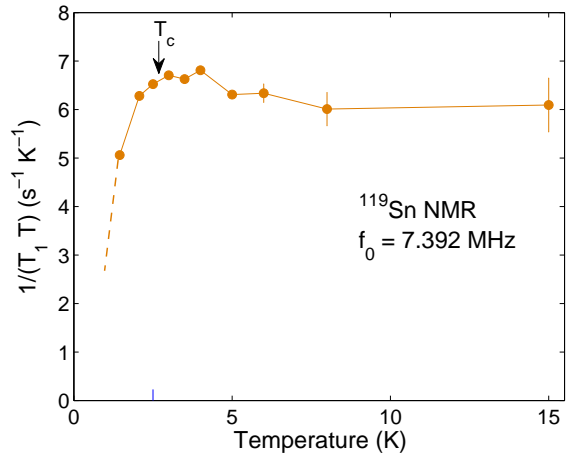}
\caption{\label{Fig:NMR-T1T}(Color online) ${}^{119}$Sn spin-lattice relaxation rate divided by temperature, $1/(T_1 T)$, vs.\ temperature measured at $\mu_0H_{ext} = 0.463$ T. The product is mostly constant in the normal state, but it drops for $T < 2.5$ K, reflecting the onset of superconductivity.}
\end{figure}

The magnetization recovery curves can be fitted using a single-exponential model with increasing $T_1$ as the temperature decreases. The relaxation times are of the order of tens of milliseconds, typical of the efficient electron relaxation mechanisms of metallic compounds.
Figure~\ref{Fig:NMR-T1T} plots the inverse of the $T_1 T$  as a function of $T$. In the normal state we obtain a nearly temperature-independent $T_1 T$ product, corresponding to the Korringa prediction of a linear temperature dependence of $1/T_1$ for electronic relaxation processes.\cite{korringa50}
The steep decrease in $1/(T_1 T)$ below $T_c$ (2.5 K at 0.463 T) reflects the decreasing number of electrons available for relaxation processes, due to the formation of the singlet pairs in the superconducting state and the opening of the relative gap.

\begin{figure}[th]
\centering
\includegraphics[width=0.42\textwidth]{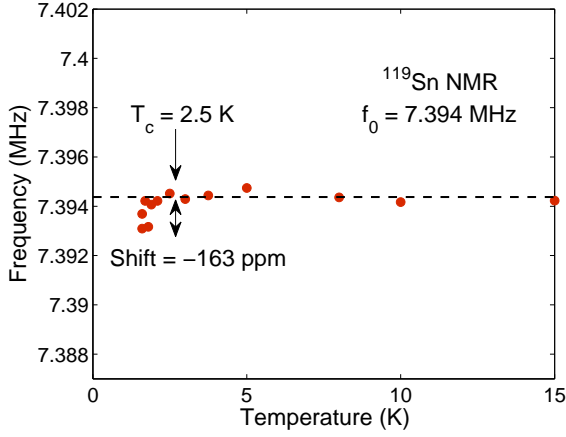}
\caption{\label{Fig:freq-vs-T}(Color online) Temperature dependence of the first moment of the $^{119}$Sn NMR resonance line.}
\end{figure}

Figure~\ref{Fig:freq-vs-T} shows the temperature dependence of the first moment of the NMR resonance line. The line position was calibrated via a secondary Al reference sample.\cite{harris01} In the normal state we detect a Knight shift $K = 1000(50)$ ppm,
to be compared with a shift of about 7000 ppm (isotropic part) measured in metallic tin.\cite{borsa66}

Between the normal and the superconducting state
we observe an additional diamagnetic shift of 163 ppm,
in agreement with the reduction of the spin susceptibility for singlet-spin pairing. The
exiguity of the shift can be explained by the relatively large spin-orbit coupling
which characterizes heavy atoms such as tin ($Z=50$).
Indeed, in metallic Sn it has been found that spin-orbit interaction destroys the spin as a good quantum number and leads to a residual spin
susceptibility at $T=0$ which is $\sim 80\%$ that of the normal state susceptibility.\cite{maclaughlin76} In Y\-Pd$_2$\-Sn we expect a similar effect.
By using the electronic parameters reported in Sec.~\ref{Ssec:parameters} we estimate a reduction of spin susceptibility in the superconducting state
$\chi_s / \chi_n \simeq 1 - 2 \Delta(0) \tau_{so} / \hbar \simeq 0.75$. At $T = 2$ K
this may lead to a $\sim 100$ ppm drop of the normal-state Knight-shift, in agreement with our observations.

The small shifts of the ${}^{119}$Sn and ${}^{89}$Y nuclei ($K_{89} = 2000$ ppm at room temperature)\cite{hoting12} are consistent with band structure calculations, which show that the conduction electrons at the Fermi energy $E_{\mathrm{F}}$ are mainly composed of 4$d$ electrons derived from Pd atoms, indicating that in Y\-Pd$_2$\-Sn it is the Pd sublattice which is mainly responsible for the superconductivity.
The fact that 4$d$ and other bands cross the Fermi energy may also explain why the value of $1 / (T_1 T K^2)$ is higher than that expected from the Korringa ratio for simple single-band systems $4 \pi k_{\mathrm{B}} \gamma_{\mathrm{Sn}}^2 / (\hbar \gamma_{e}^2)$.

\
\

\subsection{\label{Ssec:ZF mSR} Zero-field \mSR}

We investigated the magnetic response of Y\-Pd$_2$\-Sn by means of ZF-$\mu$SR. Elemental metallic palladium is on the verge of FM, \cite{parolin07} as shown by its large temperature-dependent paramagnetic susceptibility and by the fact that it sustains over-damped collective magnetic excitations or paramagnons.\cite{doubble10} The Fermi level of metallic Pd lies just above a large peak of its DOS, at the top of the $d$ bands. In Y\-Pd$_2$\-Sn the density of states at $E_{\mathrm{F}}$ is very close to that of elemental fcc Pd,\cite{klimczuk12} which can be easily tuned towards an itinerant ferromagnetic ground state either by introducing dilute magnetic impurities or by expanding its lattice. Several calculations have shown that a volume increase by 20\% of the fcc lattice or a change of atomic coordination can lead to the formation of magnetic moments of $\sim 0.2$~$\mu_{\mathrm{B}}$.\cite{moruzzi89,huger05} In \ypd the (cubic) Pd sublattice, which is responsible for the superconductivity, is expanded with respect to elemental Pd.
Previous reports of weak magnetism in Y\-Pd$_2$\-Sn,\cite{amato04}  as well as $\mu$SR and inelastic neutron scattering results in the Yb\-Pd$_2$\-Sn compound indicating a possible coupling of the magnetic fluctuations with superconductivity \cite{aoki00} are intriguing. These results raise the question about the role of Pd and the possible presence of a subtle interplay between its nearly FM character and the  superconductivity.

One of the most sensitive and direct ways of detecting very weak magnetism is muon spin rotation in zero field. Any sign of magnetism related to Pd, which may be enhanced at $T_c$, due to electronic instabilities or to paramagnons, should increase the muon spin relaxation rate.
We performed ZF-$\mu$SR measurements as a function of temperature after carefully compensating the earth magnetic field.
Several fit functions were used in search of signatures of weak electronic moments. The $\mu$SR spectra could be best fitted by the so called Kubo-Toyabe function \cite{yaouanc11} $G(t)= \frac{1}{3} + (1- \Delta^2 t^2) e^{-\frac {\Delta^2 t^2} {2}}$.  The variance of the corresponding Gaussian field distribution $\Delta^2 / \gamma_{\mu}^2$ and the static character shown by the decoupling behavior in weak longitudinal fields are consistent with a muon spin relaxation solely due to static nuclear moments. The relaxation is nearly constant in the normal and superconducting state. These results exclude the presence of static or dynamic electronic magnetism, which may be associated with Pd or the superconducting state.

\begin{figure}[th]
\centering
\includegraphics[width=0.42\textwidth]{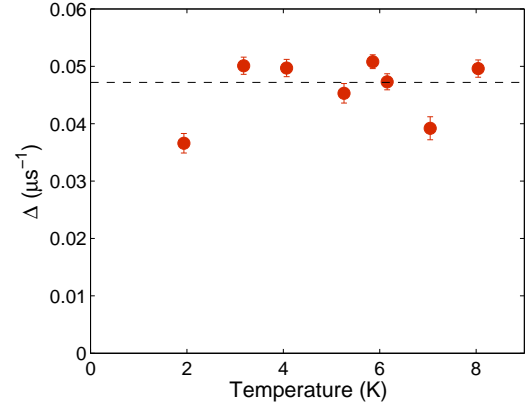}
\caption{\label{Fig:ZF-KT-fit}(Color online) Temperature dependence of the ZF-$\mu$SR relaxation rate.}
\end{figure}
\

\subsection{\label{Ssec:parameters} Normal and superconducting state parameters}

From the data obtained so far we can estimate various normal and superconducting state parameters of  Y\-Pd$_2$\-Sn.
From the measured $B_{c2}$(0)
we obtain the Ginzburg-Landau coherence length $\xi_{\mathrm{GL}} = \sqrt{\frac {\Phi_0}{2\pi B_{c2}}} \cong $ 23 nm, where $\Phi_0 = 2.07 \times 10^{-15}$ T\,m$^2$ is the quantum of magnetic flux. With $\lambda(0)=\lambda_{\mathrm{GL}}$ = 212 nm this gives a Ginzburg-Landau parameter for this type-II superconductor $\kappa=\frac{\xi_{GL}}{\lambda_{GL}} \approx$ 9.
Using the values $\lambda_{\mathrm{GL}}$ and $\mu_0 H_{c2}(0)$ the first critical field is estimated
from $B_{c1}(T)=\frac{\Phi_0}{4 \pi \lambda_{\mathrm{GL}}^2}(\ln \kappa + 0.5)$ to be 9.9 mT, in excellent agreement with field-dependent magnetization data $M(H)$.\cite{klimczuk12}

From our measurement of $\Delta$(0) we determine the BCS coherence length $\xi_0=  \hbar^2 (3 \pi^2 n_e)^{\frac{1}{3}} / [m (1+\lambda_{ep}) \pi \Delta(0)]$ (where $\lambda_{ep}=0.7$ is the electron-phonon coupling obtained by solving the McMillan expression \cite{mcmillan68} for $\lambda_{ep}$ with our \Tc = 5.4 K and a Debye temperature \cite{klimczuk12} $\Theta_{\mathrm{D}}= 210$ K). The electron density $n_e \simeq 3.3 \times 10^{23}$ cm$^{-3}$ was estimated from calculations of the density of states.\cite{winterlik09} This value corresponds to estimates based on a free-electron model for a system of 27 valence electrons, such as Y\-Pd$_2$\-Sn,\cite{klimczuk12} and provides $\xi_0 = 361$ nm.

Similarly, using the value of the residual resistivity above $T_c$ ($\rho_n = 9.2$ m$\Omega\,$cm) we could estimate the mean free path $\ell= \hbar (3 \pi^2)^{\frac{1}{3}}/ (n_e^{\frac{2}{3}} \rho_n e^2)$ = 2.9 nm.
It should be noted that very similar values for $\xi_{\mathrm{GL}}$, $\xi_{0}$, and $\ell$ are obtained using the expressions based on the evaluation of the superconducting and normal state parameters from the BCS-Gorkov equations reported in Ref.~[\onlinecite{orlando79}], where these quantities are expressed in terms of $T_c$, $\rho_n$, the Fermi surface $S$, and $\gamma_n$, the normal-state electronic specific heat coefficient.\cite{note1}
Moreover, the good agreement of $\xi_{\mathrm{GL}}$ determined above with the value obtained from $\sqrt{\xi_0 \ell} \simeq 32$ nm,
gives us further confidence about the correct estimate of these parameters.

The low value of $\ell / \xi_0 = 0.008$ means that \ypd is a superconductor in the very dirty limit, but remains a good metal with $k_{\mathrm{F}} \ell \simeq 60$.
From our determination of $\lambda$, which represents the effective magnetic penetration depth, we can estimate a maximum value of the superfluid density $n_s = \frac{m(1+\lambda_{ep})}{\mu_0 e^2 \lambda^2} \simeq 1 \times 10^{23}$ cm$^{-3}$, which, for such a dirty superconductor, is surprisingly close to $n_e$.

A striking aspect of Y\-Pd$_2$\-Sn, in view of its character of conventional $s$-wave superconductor, is the very large variation of
the values of $T_c$ reported in the literature. A study of phase equilibria and superconductivity in systems containing Pd, Sn, and Y has shown that a relative variation of the nominal concentration of a few percent leads to changes of $T_c$ from 2.7 K to 5.5 K. A value of 5.4 K, as in our case (to be compared with the generally reported value \cite{klimczuk12} of 4.7 K), has been found for a composition close to the ideal concentration ratio.\cite{jorda85} The high $T_c$ value in our case is significant because, in contrast to resistive transitions, which can reflect the presence of isolated superconducting paths, it is also observed by the volume sensitive $\mu$SR, indicating a full \scg fraction. It is remarkable that the reported variations of \Tc within different samples are even more pronounced than those obtained with most of the magnetic rare earths (Y$_{1-x}$\-RE$_{x}$\-Pd$_{2}$\-Sn) and only comparable with the substitution of Y with Gd.\cite{malik85}
A sensitive dependence of $T_c$ on disorder has been observed in similar Heusler compounds\cite{winterlik09} and also band structure calculations show that already small amounts of disorder may produce distinct changes in the electronic structure and the related properties, reflecting the presence of sharp features in the electronic density of states.
We recall that, due to the proximity of the atomic numbers of palladium (46) and tin (50) and their similar nuclear scattering lengths,
techniques such as X-ray and neutron diffraction are fairly insensitive to disorder. Therefore, partial disorder in the palladium and tin sites is quite possible
and may play a role beyond the Anderson theorem, making the
significant scattering of critical temperatures easily 
related to the very dirty character of this superconductor.

\section{\label{Sec:Conclusions}Conclusions}

\ypd is found to be a very dirty bulk superconductor with $T_c = 5.4$ K.
\mSR spectra in the vortex state can be well fitted with a superconducting $s$-wave
function having a superconducting gap of 0.85(3) meV. We find that the \mSR data are consistently reproduced in a broad range of applied fields
with a field-dependent superconducting gap, which has a similar dependence on field 
as the critical temperature. The ratio $2\Delta(0) / (k_{\mathrm{B}} T_c) = 3.85$
is consistent with the presence of an important electron-phonon coupling in this compound.
The effective magnetic penetration depth, as resulting from fits of $\mu$SR data, is $\lambda$(0) = 212(1) nm. Although NMR on ${}^{119}$Sn appears of limited effectivity in determining the gap parameter in the superconducting state, it nonetheless detects the spectral changes of the linewidth related to the presence of superconducting vortices, in good agreement with the numerical values obtained from the $\mu$SR  data analysis.
The temperature dependence of the NMR Knight shift and that of the relaxation rate $1 / T_1$ both reflect the opening
of the superconducting gap. Despite the high sensitivity of the ZF-$\mu$SR technique, we find no indications of static or dynamic magnetism associated with Pd in this compound.

\begin{acknowledgments}
H.S.\ and E.M.\ acknowledge support from the NCCR program MaNEP. T.S.\ and M.P.\ acknowledge support from the Schweizer Nationalfonds (SNF) and the NCCR program MaNEP. We thank A.\ Yaouanc, R.F.\ Kiefl, A.\ Maisuradze, and H.-R.\ Ott for useful discussions.
\end{acknowledgments}


\begin{thebibliography}{99}
\bibitem{note1} Here we take $S$ from the free-electron
theory and use the $\gamma_n$ value reported in Ref. [\onlinecite{klimczuk12}].
\end{thebibliography}
\end{document}